\documentclass[aps,prl,twocolumn,floatfix]{revtex4}
\usepackage{graphicx}
\usepackage{amsmath}
\begin{document}

\title{Nuclear spin gyroscope based on an atomic co-magnetometer}
\author{T. W. Kornack, R. K. Ghosh and M. V. Romalis}
\affiliation{Department of Physics, Princeton University,
Princeton, NJ 08550 USA}
\begin{abstract}
We describe a nuclear spin gyroscope based on an
alkali-metal--noble-gas co-magnetometer. An optically pumped
alkali-metal vapor is used to polarize the noble gas atoms and
detect their gyroscopic precession. Spin precession due to
magnetic fields as well as their gradients and transients can be
cancelled in this arrangement. The sensitivity is enhanced by
using a high-density alkali-metal vapor in a spin-exchange
relaxation free (SERF) regime.  With a K--$^3$He co-magnetometer
we demonstrate rotation sensitivity of $5 \times 10^{-7} {\rm
rad/sec/Hz^{1/2}}$. The rotation signal can be increased by a
factor of 10 using $^{21}$Ne due to its smaller magnetic moment
and the fundamental rotation sensitivity limit for a $^{21}$Ne
gyroscope with a 10 cm$^{3}$ measurement volume is about $2\times
10^{-10}$ rad/sec/Hz$^{1/2}$.

%This co-magnetometer is also a promising tool in searches for Lorentz violation,
      %spin-dependent forces and other interactions beyond the Standard Model.
\end{abstract}

\maketitle

Sensitive gyroscopes find a wide range of applications, from
inertial navigation to studies of Earth rotation and tests of
general relativity \cite{Stedman1997}. A variety of physical
principles have been utilized for rotation sensing, including
mechanical sensing, the Sagnac effect for photons
\cite{Stedman1997,FiberOpRev} and atoms
\cite{AtomIntPRL,ColdAtomInt}, the Josephson effect in superfluid
$^4$He and $^3$He \cite{Avenel} and nuclear spin precession
\cite{Woodman87}. While state-of-the-art mechanical gyroscopes,
such as those developed for Gravity Probe B \cite{GravityB},
remain unchallenged in terms of sensitivity, their extremely high
cost and difficulty of fabrication motivate the development of
simpler, smaller and more robust rotation sensors.

Here we describe a new gyroscope based on nuclear spin precession.
Unlike the atom and photon interferometric gyroscopes based on the
Sagnac effect, nuclear spin gyroscopes do not require a large area
enclosed by the interferometer and can be made quite compact.
Previous nuclear spin gyroscopes \cite{Woodman87} have suffered
from high sensitivity to magnetic fields. We show that a
co-magnetometer using spin-polarized noble gas and alkali-metal
vapor can eliminate the sensitivity to magnetic fields, their
gradients and transients. High short-term rotation sensitivity can
be achieved with an alkali-metal magnetometer operating in the
SERF regime \cite{Allred}. For example, magnetic field sensitivity
of 0.5 fT/Hz$^{1/2}$ that has been demonstrated in a K
magnetometer \cite{Kominis} would result in a rotation sensitivity
of $1 \times 10^{-8}$ rad/s/Hz$^{1/2}$ in a K-$^{21}$Ne gyroscope.
The bandwidth and transient response of the gyroscope are also
significantly improved compared with earlier spin gyroscopes by
damping due to coupling between noble gas and alkali-metal spins.
We describe an experimental implementation of the gyroscope using
K and ${^3}$He atoms and demonstrate short term rotation
sensitivity of  $5 \times 10^{-7}$ ${\rm rad/sec/Hz^{1/2}}$ with a
sensing volume of only 0.5 cm$^{3}$.   We also present a
theoretical analysis and experimental measurements of the
gyroscope response to various perturbations, and derive
fundamental limits for its performance.

The co-magnetometer consists of a spherical glass cell containing
an alkali metal, several atmospheres of noble gas and a small
quantity of nitrogen. Alkali atoms are polarized by optical
pumping and transfer the polarization to the noble gas nuclei by
spin-exchange collisions. A probe laser passes through the cell
perpendicular to the pump laser and measures the direction of the
alkali-metal polarization, which is strongly coupled to the
nuclear polarization of the noble gas due to the imaginary part of
the spin-exchange cross-section. For sufficiently high buffer gas
pressure in a spherical cell, this coupling can be represented by
an effective magnetic field that one spin species experiences from
the average magnetization of the other, $\mathbf{B} = \lambda
\mathbf{M}$, where $\lambda = 8 \pi \kappa_0 / 3$ \cite{Shaefer}.
Here $\kappa_0$ is an enhancement factor due to the attraction of
the electron wavefunction towards the noble gas nucleus and ranges
from about 5 to 600 for different alkali-metal--noble-gas pairs
\cite{Walker}.

It was shown in  \cite{Kornack02} that the co-magnetometer is
accurately described by a system of coupled Bloch equations for
the electron and nuclear polarizations, $\mathbf{P^e}$ and
$\mathbf{P^n}$:
\begin{align}
\frac{\partial \mathbf{P^e}}{\partial t} &= \mathbf{\Omega} \times
\mathbf{P^e}+
 \frac{\gamma_e}{Q(P^e)}(\mathbf{B}+\lambda M^{n}_0 \mathbf{P^n} + \mathbf{L} ) \times
 \mathbf{P^e} \nonumber \\
&+(R_{p} \mathbf{s_{p}}  + R^{e}_{se}\mathbf{P^n} +R_{m} \mathbf{s_{m}} - R_{tot}\mathbf{P^e} )/Q(P^e) \nonumber \\
\frac{\partial \mathbf{P^n}}{\partial t} &= \mathbf{\Omega} \times
\mathbf{P^n}+\gamma_n (\mathbf{B}+\lambda M^{e}_{0} \mathbf{P^e} )
\times
\mathbf{P^n}  \nonumber \\
&+ R^n_{se}(\mathbf{ P^e} - \mathbf{P^n} )-R^n_{sd} \mathbf{P^n}
\label{Bloch}
\end{align}
Here $\mathbf{\Omega}$ is the mechanical rotation, $\gamma_e = g_s
\mu_B / \hbar$ and $\gamma_n = \mu_{n} /I \hbar$ are the
gyromagnetic ratios of electron and nuclear spins. $M^e_0$ and
$M^n_0$ are the magnetizations of electron and nuclear spins
corresponding to full spin polarizations. $\mathbf{L}$ is the
effective magnetic field for alkali-metal spins created by the
light shift from pumping and probing lasers \cite{HapperLS}. $R_p$
and $R_m$ are the pumping rates of the pump and probe laser beams
while $\mathbf{s}_p$ and $\mathbf{s}_m$ give the directions and
magnitudes of their photon spin polarizations. $R^{e}_{se}$ is the
alkali-metal--noble-gas spin-exchange rate for an alkali atom and
$R^{n}_{se}$ is the same rate for a noble gas atom. $R_{tot}$ is
the total spin relaxation rate for alkali atoms; $R_{tot} = R_p +
R_m + R^e_{se} + R^e_{sd}$, where $R^e_{sd}$ is the electron spin
destruction rate. $R^n_{sd}$ is the nuclear spin relaxation rate.
$Q(P^e)$ is the electron slowing-down factor due to hyperfine
interaction and spin-exchange collisions \cite{Savukov}. For
alkali metal isotopes with $I=3/2$ in the regime of fast
alkali-metal spin-exchange,  $Q(P^e)$ ranges from 6 for low $P^e$
to 4 for $P^e \approx 1$.

The co-magnetometer is nominally configured with the pump beam
directed along the $\hat{z}$ axis and the probe beam directed
along the $\hat{x}$ axis. A compensating field $\mathbf{B} =
B_{c}\hat{z} = -(B^n+B^e) \hat{z}$ exactly cancels the field due
to the magnetized atoms \cite{Kornack02}. Here the effective field
from nuclear magnetization $B^n =\lambda M^{n}_{0} P^n_z$ is
typically on the order of a few mG and the effective field from
the electron magnetization $B^e =\lambda M^{e}_{0} P^e_z$ is on
the order of a few $\mu$G. The light shifts can be set to zero,
$\mathbf{L}=0$, because the pump beam is tuned to the center of
the optical resonance and the probe beam is linearly polarized.
Under these conditions the gyroscope signal, proportional to the
optical rotation of the probe beam  due to $P^e_x$, is accurately
given by
\begin{equation}
S = \frac{\gamma_e \Omega_y  P^e_z }{\gamma_n R_{tot}} \left(1
-\frac{\gamma_n}{\gamma_e} Q(P^{e}) - C^n_{se}\right) \label{rot}
\end{equation}
Thus, the signal is proportional to  rotation about the $\hat{y}$
axis and is enhanced by the ratio $\gamma_e/\gamma_n \gg 1$. The
nuclear spin-exchange correction factor $C^n_{se}= (\gamma_e P^e_z
R^n_{se})/(\gamma_n P^n_z R_{tot})$ is typically on the order of
$10^{-3}$.

\begin{figure}
\includegraphics[width=0.85\linewidth]{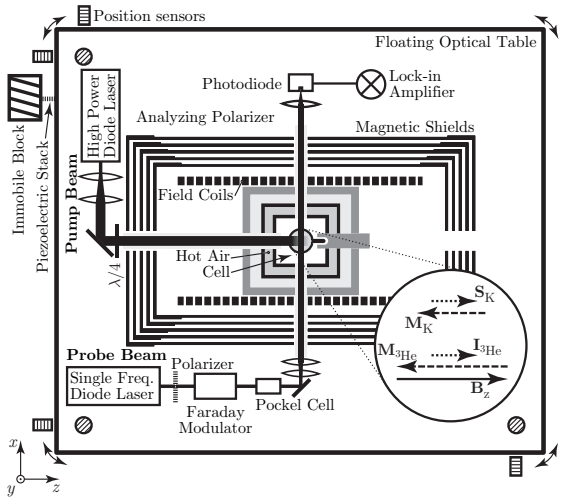}
\caption{Schematic of the experimental setup, consisting of a 2.5
cm diameter aluminosilicate glass cell containing K and $^3$He
atoms heated to 170${^\circ}$C in an oven inside magnetic shields.
Diode lasers are used for optical pumping and probing of K vapor.
The floating optical table is rotated with a piezo stack and
monitored by non-contact position sensors.} \label{GyroExperiment}
\end{figure}

Our experimental implementation of the gyroscope using K and
$^3$He atoms is similar to the setup in \cite{Kornack02} and is
depicted in Fig. \ref{GyroExperiment}. The floating optical table
is equipped with a piezo actuator to induce small rotations and 6
non-contact position sensors to measure the resulting rotational
motion. Feedback circuits were implemented to control the
wavelength and intensity of pump and probe lasers. Magnetic fields
and light shifts were periodically zeroed using a modulation
procedure described below.

\begin{figure}
\includegraphics[width=0.9\linewidth]{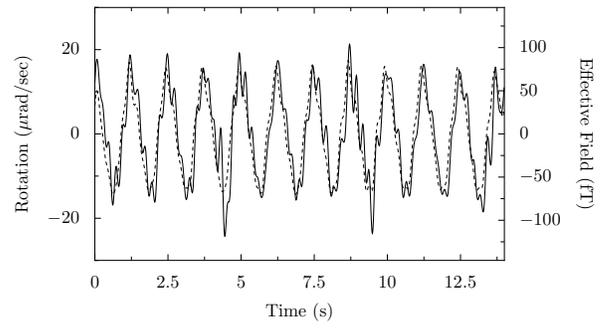}
\caption{Angular velocity due to a piezo excitation measured with
the co-magnetometer (solid line) and position sensors (dashed
line), plotted with no free parameters.}
\label{GyroRotationSignal}
\end{figure}

Fig. \ref{GyroRotationSignal} shows the angular velocity signal
measured by the spin gyroscope compared with the angular velocity
$\Omega_y$ obtained from the position sensors. The gyroscope
sensitivity was calibrated as described below and agreed with
mechanical   measurements within the calibration accuracy of 3\%.
We also verified that the gyroscope is insensitive to the other
two components of angular velocity.

\begin{figure}
\includegraphics[width=0.9\linewidth]{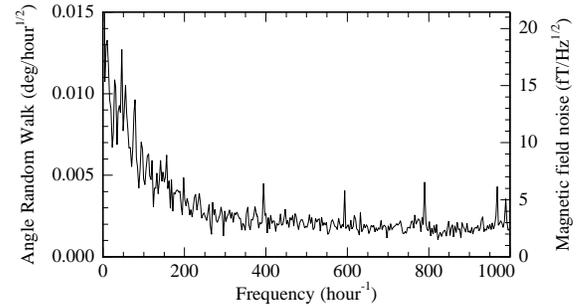}
\caption{Fourier spectrum of the gyroscope rotation noise.
Discrete peaks are an artifact of periodic zeroing of the $B_z$
field. The $1/f$ noise knee is at about 0.05 Hz.} \label{spectrum}
\end{figure}

The sensitivity of the gyroscope is shown in Fig. \ref{spectrum}.
The angle random walk (ARW) is 0.002 deg/hour$^{1/2}$ or $5 \times
10^{-7}$ rad/s/Hz$^{1/2}$ in the white noise region and
corresponds to a magnetic field sensitivity of 2.5 fT/Hz$^{1/2}$.
The low frequency angle drift of the gyroscope in the present
implementation is about 0.04 deg/hour.

To understand the effects of various experimental imperfections it
is important to consider small corrections to Eq. (\ref{rot}). The
only first order dependence on the magnetic fields or light-shift
fields comes from the $B_x$ field:
\begin{equation}
 S(B_x)=B_x P^e_z (C^e_{se} +C^n_{se})/B^n,
\end{equation}
where the electron spin-exchange correction $C^e_{se}=(R^e_{se}
P^n_z)/(R_{tot} P^e_z)$ and is on the order of $10^{-2}$. Because
electron and nuclear spin-exchange corrections are small and $
R_{tot} \ll \gamma_e B^n$,  the field sensitivity is suppressed by
a factor of $10^5$ under typical conditions. Misalignment of the
pump and probe beams by an angle $\alpha$ away from 90$^\circ$
gives a signal $S = \alpha R_{p}/R_{tot}$. For typical conditions,
1 $\mu$rad of misalignment gives a false rotation of $10^{-8}$
rad/sec. Misalignment can be distinguished from a true rotation
signal by its dependence on the pumping rate $R_p$. Possible
circular polarization of the probe laser $s_m$ also gives a
first-order signal $S = s_m R_{m} / R_{tot}$ but it can be  set to
zero by minimizing the light shift as described below.

Other imperfections only affect the signal to second order in
small quantities. For example, the signal due to the $B_y$ field
is given by
\begin{equation}
S(B_y)=\frac{ \gamma_e B_y P^e_z}{B^n R_{tot}} \left(B_z -
(B_z+L_z) C^e_{se}-(2 B_z+L_z) C^n_{se}\right)
\end{equation}
where $B_z$ is a small detuning away from the compensation field
$B_c$. In addition to suppressing imperfections by two small
factors, such second order dependence  allows us to calibrate the
co-magnetometer and provides a mechanism for zeroing many offsets.
For example, to set  $B_z$  to zero we apply a modulation to the
$B_y$ field, measure the response as a function of $B_z$ and find
the point where it vanishes.  The slope of the response is given
by
\begin{equation}
\frac{\partial^2 S}{\partial B_y \partial B_z} =\frac{ \gamma_e
P^e_z}{B^n R_{tot}} (1-C^e_{se}-2 C^n_{se}) \simeq \frac{ \gamma_e
P^e_z}{|B_c| R_{tot}} \label{Bymodslope}
\end{equation}
The approximations in the last step are accurate to better than
1\%  because under typical conditions  $B^e \ll B^n$ and
$C^e_{se},C^n_{se}\ll 1$. The measurement of the slope gives a
calibration of the gyroscope signal (\ref{rot}) in terms of the
known applied magnetic fields $B_y$, $B_z$ and $B_c$. Most other
field, light shift and alignment imperfections can be minimized in
a similar way with an appropriate choice of modulation. For
example, a term in the signal proportional to $L_x L_z$ allows us
to minimize the light shifts of the pump and probe beams by
modulating one of them and adjusting the other to get zero
response. Since $L_x \propto s_m$, this also minimizes the probe
circular polarization.

\begin{figure}
\includegraphics[width=0.9\linewidth]{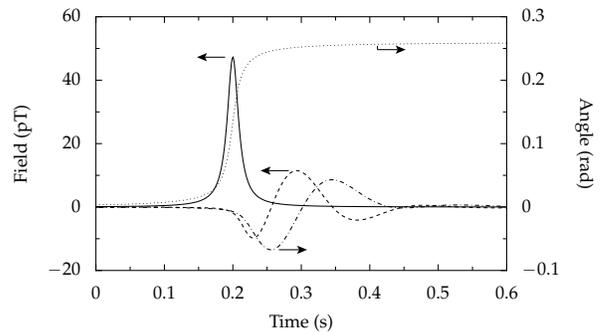}
\caption{Response of the co-magnetometer (dashed line)  to a
magnetic field transient (solid line), plotted against the left
axis. The gyroscope rotation angle (dash-dot line), proportional
to the integral of the co-magnetometer signal, is much smaller
than the expected rotation angle for an uncompensated K
magnetometer (dotted line), plotted against the right axis. }
\label{gyrotransientplot}
\end{figure}

\begin{figure}
\includegraphics[width=0.9\linewidth]{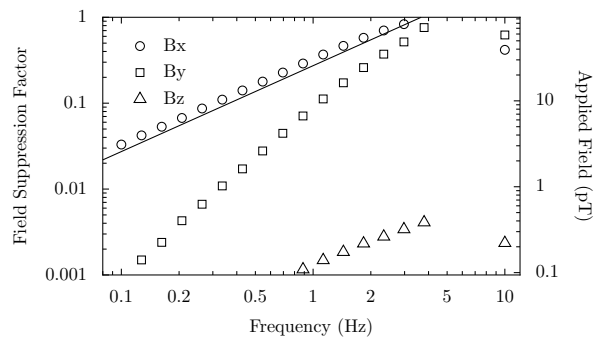}
\caption{Response of the gyroscope to uniform oscillating fields
created by coils inside the shields in the $\hat{x}$, $\hat{y}$
and $\hat{z}$ directions as a function of frequency. The gyroscope
signal is suppressed in comparison to the signal expected from an
uncompensated alkali-metal magnetometer $S=\gamma_e B
P^e_z/R_{tot}$. The field sensitivity is highest in the $\hat{x}$
direction and agrees with Eq. (\ref{suppr}) (solid line).}
\label{fieldsuppression}
\end{figure}

The transient response of the gyroscope is also improved in the
co-magnetometer configuration. In navigation applications, the
rotation frequency is integrated over time to obtain the rotation
angle. Using the Green's function  for linearized Bloch equations
\cite{Kornack02}, it can be shown that the integral of the signal
is proportional to the total angle of mechanical rotation about
the $\hat{y}$ axis independent of the time dependence of
$\Omega_y$. Furthermore, the net rotation angle generated by an
arbitrary magnetic field transient is equal to zero as long as
spin polarizations are rotated by a small angle during the
transient. Fig. \ref{gyrotransientplot} shows the response of the
gyroscope to a transient magnetic field spike, demonstrating
reduction of the spin rotation angle by a factor of 400 relative
to an uncompensated K magnetometer. For an oscillating magnetic
field the largest contribution to the signal comes from the
$B_{x}$ field. For $B_{x}=B_0 \cos(\omega t)$ the signal is equal
to
\begin{equation}
S(\omega)= \frac{ B_0 \gamma_e P^e_z \omega \sin(\omega
t)}{\gamma_n B^n R_{tot}} \label{suppr}
\end{equation}
and is suppressed at low frequencies. The response of the
co-magnetometer to oscillating magnetic fields is shown in Fig.
\ref{fieldsuppression} and is in close agreement with Eq.
(\ref{suppr}),

The coherent spin coupling between alkali-metal and nuclear spins
also causes fast damping of transient excitations, which decay
with a time constant
\begin{equation}
\frac{1}{T_d} = \frac{ \gamma_n \gamma_e B^n B^e
R_{tot}}{(\gamma_n  B^n Q(P^e))^2+R_{tot}^2}
\end{equation}
The decay time $T_d$ is on the order of 0.1 sec and is not limited
by the nuclear spin relaxation time $(R^n_{sd})^{-1}$ that is
typically thousands of seconds. Fast damping of nuclear spin
coherence ensures that coherent nuclear quadrupole interactions
with cell walls, which produce beats in spin precession signal for
isotopes with $I>1/2$ on a time scale $(R^n_{sd})^{-1}$
\cite{HapperQuad,Chupp} do not cause a significant effect.

The co-magnetometer also effectively suppresses magnetic field
gradients even though alkali-metal and noble gas polarizations
have somewhat different spatial distributions. The degree of
nuclear polarization is very uniform across the cell because the
rate of diffusion $R_D=D/a^2$, where $D$ is the diffusion constant
and $a$ is the radius of the cell, is much faster than $R^n_{sd}$.
The direction of nuclear polarization is parallel to the local
magnetic field as long as the nuclear spin precession frequency
$\gamma_n B^n \gg R_D$ \cite{Cates}. Thus, nuclear magnetization
largely cancels the non-uniform external field point-by-point. The
limits of this cancellation are determined by the degree of
nuclear spin misalignment given by $R_D/\gamma_n B^n$ and the
local variation in $B_c$ due to non-uniform alkali-metal
polarization, on the order of $B^e/B^n$. Both effects are on the
order of $10^{-3}$ under typical conditions.  We measured the
sensitivity to first order magnetic field gradients using internal
gradient coils. The quasi-static signals from gradient fields are
suppressed relative to $S_g=\gamma_e |\nabla B| a P^e_z/R_{tot}$
by a factor of 500 to 5000.

The fundamental limit on gyroscope sensitivity is due to spin
projection noise. We performed a quantum trajectory simulation of
the coupled spin system (\ref{Bloch}) to show that for most
parameters the measurement uncertainty is dominated by the
alkali-metal spins. The rotational uncertainty per unit bandwidth
is given by $\delta \Omega_y = (\gamma_n/\gamma_e) [Q(P^e)
R_{tot}/n V]^{1/2}$ where $n$ is the density of alkali-metal atoms
and $V$ is the measurement volume. $^{21}$Ne gives the best
fundamental sensitivity and suppression of systematic effects
because it has a small gyromagnetic ratio $\gamma_n$, ten times
smaller than  $^3$He. Using the  K-$^{21}$Ne spin relaxation
cross-section measured in \cite{Franz} we estimate the fundamental
sensitivity to be $2 \times 10^{-10}$ rad/sec/Hz$^{1/2}$ for a
measurement volume of 10 cm$^{3}$, K density of 10$^{14}$
cm$^{-3}$ and $^{21}$Ne density of $3\times 10^{20}$ cm$^{-3}$.
Detection of off-resonant optical rotation allows one to approach
the spin projection noise even with imperfect detectors by making
a quantum-non-demolition measurement of the alkali-metal spin in
an optically-thick vapor.

For comparison, gyroscopes utilizing the Sagnac effect have
achieved sensitivities of $2 \times 10^{-10}$ rad/sec/Hz$^{1/2}$
using a ring laser with an enclosed area of 1 m$^2$
\cite{Stedman2003} and $6 \times 10^{-10}$ rad/sec/Hz$^{1/2}$
using an atomic inteferometer with a path length of 2 m
\cite{AtomIntSens}. More compact atomic inteferometers using cold
atoms that are presently being developed have a projected
shot-noise sensitivity of $3 \times 10^{-8}$ rad/sec/Hz$^{1/2}$
\cite{ColdAtomInt} and $2 \times 10^{-9}$ rad/sec/Hz$^{1/2}$
\cite{AtomGermany}. Compact state-of-the-art fiber-optic
gyroscopes have a reported sensitivity of $2 \times 10^{-8}$
rad/sec/Hz$^{1/2}$ \cite{Honeywell}. Thus, the gyroscope described
here is promising as a compact rotation sensor that can rival
existing technologies. It's relative simplicity makes it amenable
to miniaturization with techniques  developed for compact atomic
clocks \cite{Kitching}. Many aspects of the system, such as
magnetic shielding and mechanical stability will improve with
smaller size. Small size and fast transient response may also
allow reduction of the gyroscope long-term drifts using active
rotation techniques \cite{Subnav}.

In conclusion, we have described  the operation and performance of
a K--$^3$He co-magnetometer gyroscope. It has a high short term
sensitivity with a small measurement volume and is insensitive to
external perturbations. Further improvement is possible by
switching to $^{21}$Ne gas and improving the sensitivity of
optical rotation measurements at low frequencies to approach the
spin-projection noise.  We thank Tom Jackson, Igor Savukov,
Charles Sule and Saee Paliwal for assistance in the lab. This work
was supported by NASA, NSF, a NIST Precision Measurement grant,
and the Packard Foundation.
%The development of this co-magnetometer is motivated by
% experimental searches for CPT violation and nuclear EDMs.
%Rotation is just one example of an interaction whose
%coupling to electron and nuclear spins are not proportional
%to their magnetic moments; other examples include Lorentz/CPT
%violation, a permanent electric dipole moment and
%spin-dependent forces. If the present drifts can be reduced,
%we expect to achieve sensitivities to CPT-violating fields that
%surpass existing limits.

\bibliographystyle{unsrt}

\end{document}